\documentclass[showpacs,preprintnumbers,amsmath,amssymb]{revtex4}
\usepackage{graphicx}
\usepackage{dcolumn}
\usepackage{bm}
\usepackage{epstopdf}

\begin{document}


\title{Stellar Orbit Constraints on Neutralino Annihilation at the Galactic
  Center}

\author{Jeter Hall}
 \email{jeter@physics.utah.edu}
\author{Paolo Gondolo}%
 \email{paolo@physics.utah.edu}
\affiliation{%
Department of Physics, University of Utah, 115 S 1400 E Suite 201, Salt Lake City, Utah 84112\\
}%

\date{\today}

\begin{abstract}
Dark matter annihilation has been proposed to explain the TeV gamma
rays observed from the Galactic Center.
We study constraints on this hypothesis coming from the mass profile
around the Galactic Center measured by observing stellar dynamics.
We show that for several proposed WIMP candidates, the
constraints on the dark matter
density profile from measurements of mass by infrared observations are
comparable to the constraints from the measurements of the TeV source 
extension. 
\end{abstract}

\pacs{95.35.+d, 95.85.Pw, 98.70.Rz}
\maketitle

\section{Introduction}
A significant density of Dark Matter (DM) in
the Universe has been observed on many length scales.
The first evidence of the current dark matter problem came from 
the dynamics of the Coma cluster \cite{zwicky}. Evidence of
DM on a galactic scale came from rotation curves of galaxies which
show that the orbital velocities of stars in galaxies do not follow
the mass density derived from cataloging the luminous matter\cite{babcock}.
This discrepancy can be resolved by adding a large amount of dark matter, a
DM halo, that would not be included in a count of stars, gas and dust.
More recent evidence of DM includes observations of the 
anisotropy in the cosmic microwave background, the luminosity-redshift
relation for supernovae, and the theory of Big Bang Nucleosynthesis
which
measure a baryonic density of 
$\Omega_{b} h^2 = 0.022$ and a total matter density of 
$\Omega_{m} h^2 = 0.13$ \cite{ben03,cyburt03,tonry03,spergel06}. This implies that over
80\% of the matter in the Universe is dark and non-baryonic.  
\par
One class of candidates for non-baryonic DM is the
weakly interacting massive particle, the WIMP.
Theories such as supersymmetry, an extension of the usual space-time
coordinates to
include non-commuting coordinates, naturally include WIMP candidates.
WIMPs are predicted to annihilate into
other particles with energies similar to the original mass.  The annihilation
rate is a critical parameter in determining the relic density of these
WIMPs and consequently one measure of whether they are a good candidate for
a bulk of the dark matter.
Photons will result from the annihilation, either directly, or through
pion decay or acceleration of charged annihilation products.
The annihilation could thus result in a ``WIMP star'' shining in gamma
rays with energies near the particle mass. 
\par
Recent advances in gamma-ray astronomy may allow the detection
of DM annihilation.
Ground based gamma-ray telescopes are currently sensitive to photons with
energies above 100 GeV and have reached the sensitivity of a few percent of the
Crab nebula flux.
The most sensitive
ground-based gamma-ray observatory currently in operation is the HESS
array in Namibia\cite{benbow04}.
\par
The Galactic Center (GC) has been proposed
for observations of DM annihiliation
\cite{silk87,stecker89,flores94,mayer98,bergstrom98,gondolo99,cesarini04,evans04}
because it is
close
and might have a dense concentration of dark matter resulting in a strong
signal of gamma rays.
After tentative detections of a TeV gamma-ray flux from the GC by
the VERITAS collaboration \cite{kosack04} and the CANGAROO collaboration 
\cite{tsuchiya04},
the HESS collaboration \cite{hofmann03} has initiated observations of the GC.
HESS has reported a
steady excess of TeV gamma-rays from the GC during two
observational periods of $4.7$ hours and $11.8$ hours (at the 6 sigma and 9
sigma levels respectively). 
This excess of gamma rays is confined
to a region of 3 arcminutes centered around Sagittarius A$^*$, the dynamical
center of the galaxy which is believed to host a supermassive black hole
\cite{ghez05}.
The spectrum of this excess
is a power law 
($\frac{dF}{dE} \sim E^{-\alpha}$) with
 $\alpha = 2.2 \pm 0.2$ in the energy range 
[0.2,8.8] TeV \cite{aha04}.  This spectrum is harder than typical gamma-ray
sources such as plerions and active galactic nuclei.  However, many of the
newly discovered supernova remnants have similar spectra
\cite{snaha04,snaha05}.
\par
The GC gamma-ray flux may be produced by a variety of mechanisms  
\cite{aha05bh,atoyan04,bergstrom05,horns05}. For example, the central 4 million
solar mass black hole could produce the gamma-ray flux by
accelerating electrons
in an extreme advection-dominated accretion flow,
there is a high rate of supernovae near the GC
and the shock fronts could accelerate particles to TeV energies. 
Alternatively it could be due to DM annihilation.
The GC has been suggested
as a possible site of enhanced DM annihilation~\cite{gondolo99,ullio01,merritt02,gnedin04,bertone05}  because it
has a large stellar cusp and
a million solar mass black hole in the center.
The minimum radius to which any central dark matter density features extend is
a key
unknown in predicting the gamma-ray flux from WIMP annihilation.
The
interpretation of the photon flux from the GC is not settled and could lead
at the very least to another example of an extreme particle accelerator, and
possibly could shed light on the dark matter problem.
\par
In this paper we consider the DM interpretation of the GC gamma-ray emission
and study if any more
information on this hypothesis is contained in
the dynamical measurements of the mass profile at the GC.
Advancements
in infrared astronomy are testing the small scale 
mass profile of the center of the Milky Way, down to
tens of AU.
With the W.M. Keck 10m telescope, proper 
motions of stars have been monitored near the GC since 1996 \cite{ghez98,ghez00,ghez05}.  
Entire
orbits have been or will soon be measured
around the dynamical center of the galaxy.
A strong gamma-ray signal from the GC implies a large amount of DM under
the WIMP annihilation hypothesis.  
The change in mass enclosed in a sphere with radius $d$, the distance from the
GC to the star, changes as the stars, on
highly elliptical orbits, traverse any central spherically symmetric 
density enhancement of the dark
matter.  This could lead to an observable deviation of
the orbit from a purely Keplerian orbit.
Upcoming observations will
provide direct constraints on the DM density profile in the center of the
Milky Way \cite{morris05} and help us interpret the gamma-ray flux from the GC.
\par
We use the data on the stellar orbits around 
the GC published in \cite{ghez03}.  More recent
data, including the complete orbital shapes, may 
provide further constraints \cite{morris05}.
We find that the gamma-ray flux from the GC is
compatible with annihilation of
a heavy, $\sim$ 10 TeV, DM particle with a density profile consistent with 
the stellar orbits near the GC.
Depending on the particle
physics assumptions, the stellar orbits constraint is comparable but slightly stronger than the
constraint on the source 
extension due to the angular resolution of HESS. 
Gamma-ray observations could have a very
strong signature of WIMP annihilation due to the process 
$\chi \chi \rightarrow \gamma \gamma$, 
which would create a monochromatic line in
the energy spectrum at the mass of the annihilating particle. 
Unfortunately, we find that for a TeV neutralino
the flux of the monochromatic line is too weak to be seen
with an energy resolution of 10\%, the resolution of the
atmospheric Cherenkov method.
\par
In the next Section we review the analysis used to connect the gamma-ray
emission to the stellar dynamics considered.  We define expressions for the
expected gamma-ray flux
from WIMP annihilation with the purpose of clarifying its angular dependence
and the units. Next we discuss the dark matter profiles we will
use. We then study the limits imposed on dark matter at
the GC by the astronomical mass measurements and the HESS angular
profile. Finally, in Section III we present the conclusions of the
analysis.
\section{Analysis}
The flux of photons produced by DM annihilation depends on
four factors: the annihilation products energy spectrum, the DM particle mass,
its annihilation cross-section, and the density of the DM
particles.  The energy spectrum of the annihilation products, the annihilation
cross-section, and the particle mass can be calculated once a particle model
is specified.
The density 
profile of a dark matter 
halo has long been a subject of much
debate in the literature.  Theoretical astrophysical considerations and
numerical simulations have been used to suggest a family of DM halo shapes
that could exist.  The WIMP annihilation rate is proportional to the
square of the particle density and many of the suggested DM halo shapes 
formally
diverge when the emission rate is integrated along the line of sight through
the center of the halo.  The annihilation flux in the center of the DM halo 
will dominate the flux from these divergent halos.  Astrophysically the
density profile is expected to flatten at small radii where infalling objects
can sweep out the centers of the halos through dynamical heating, although
adiabatic accretion onto central baryonic density enhancements in the centers may create dark matter
density enhancements \cite{gondolo99, ullio01, merritt02,gnedin04}.  The
annihilation rate is ultimately expected to limit the DM density
\cite{gondolo99,bertone05}.
\par
As a reference and to clarify the units of the quantities involved, we 
derive the expression of the photon flux from WIMP annihilation.
Consider a small emitting volume $dV$ at a distance $\ell$ from
a detector of collecting area $dA$ (orthogonal to the line of sight.)
This volume subtends a solid angle $d\Omega$ as seen from the detector.
Let $dN_e$ be the number of photons emitted during a time interval $dt$
from the volume $dV$.  
Assuming the emission is isotropic, a fraction
$dA/(4\pi\ell^2)$ of the emitted photons is detected.  Thus the number of
detected photons in the same amount of time $dt$ is
\begin{equation}
dN_D = dN_e \frac{dA}{4\pi\ell^2} .
\end{equation}
Specifically, for WIMP annihilation ($\chi \chi \to anything \to
\gamma$),
the number of photons emitted is
\begin{equation}
dN_e=\frac{1}{2} \frac{dN_{\gamma}}{dE} \frac{\rho^2}{m_\chi^2}
\langle\sigma v\rangle \, dE\, dt\, dV .
\end{equation}
Here $\langle\sigma v\rangle$ is the $\chi$-$\chi$ annihilation cross-section
times relative velocity, $\rho$ is the WIMP mass density, 
$m_\chi$ is the WIMP mass, and $dN_{\gamma} / dE$ is the
number of photons in the energy interval $[E,E+dE]$ produced per annihilation.
The factor of $1/2$ is there because $2$ WIMPs are required per
annihilation,  $\rho^2 \langle\sigma
v\rangle \, dt\, dV / m_\chi^2$ is the number of WIMPs annihilating and
$dN_{\gamma} / dE$ is defined per annihilation. 
The photon flux from $dV$ per unit energy at the detector then follows as
\begin{equation}
\frac{d\Phi}{dE} = \frac{dN_D}{dA\, dt\, dE} =
\frac{1}{8\pi\ell^2} \frac{dN_{\gamma}}{dE} \frac{\rho^2}{m_\chi^2}
\langle\sigma v\rangle \, dV .
\label{flux2}
\end{equation}
\par
When Eq.~(\ref{flux2}) is integrated along the line of sight,
$dV$ can conveniently be written in terms of the solid
angle $d\Omega$ as $dV = d\ell \, \ell^2 d\Omega$.  This leads to the usual
formula for the flux per unit energy per unit solid angle,
\begin{equation}
\frac{d\Phi}{dE d\Omega} = \frac{1}{8\pi} \frac{dN_{\gamma}}{dE}
\,
\frac{\langle\sigma v\rangle}{ m^{2}_{\chi}} \, \frac{dJ}{d\Omega}
\label{flux1}
\end{equation}
where
\begin{equation}\label{Jdef}
\frac{dJ}{d\Omega} = \int \rho^2 \, d\ell
\end{equation}
with the integral taken along the line of sight.
We have written the solid angle $d\Omega$ explicitly in $dJ/d\Omega$ to stress
that its units are (mass~density)$^2 \times$(length)/(solid angle), as follows
from Eq.~(\ref{flux1}) and our derivation.  This same quantity is denoted by
$J(\psi)$ in the literature, e.g. \cite{bergstrom98}.
\par
Eq.~(\ref{flux1}) can be integrated over a region ${\cal R}$ of the sky to give
\begin{equation}
\frac{d\Phi}{dE}=\int\limits_{\cal R}\frac{d\Phi}{dE d\Omega} \, d\Omega=
\frac{1}{8\pi} \frac{dN_{\gamma}}{dE} \, 
\frac{\langle\sigma v\rangle}{m^2_\chi} \, \int\limits_{\cal R}
\frac{dJ}{d\Omega} \, d\Omega
\label{intflux}
\end{equation}
When integrating over the whole
source, Eq.~(\ref{intflux}) gives a total flux of
\begin{equation}
\frac{d\Phi}{dE} = 
\frac{1}{8\pi} \frac{dN_{\chi\chi
    \to \gamma}}{dE} \frac{\langle\sigma v\rangle}{m_\chi^2}
~J
\label{flux4}
\end{equation}
where
\begin{equation}
J = \int\limits_{\rm source} \frac{dJ}{d\Omega} \, d\Omega
\end{equation}
$J$ has units of (mass~density)$^2 \times$(length). 
Several units have been used in the literature. In particular Bergstr\"om,
Ullio, and Buckley \cite{bergstrom98} used
$8.5$~kpc~$(0.3$~GeV~$c^{-2}$~cm$^{-3})^2$.
For brevity,
we introduce a Bergstr\"om-Ullio-Buckley Unit (BUBU)
\begin{equation}
1 {\rm~BUBU} = 8.5 {\rm~kpc}~(0.3 {\rm~GeV}~c^{-2}{\rm~cm}^{-3})^2 
= 2.3605 \times 10^{21} {\rm~GeV}^2~c^{-4} {\rm~cm}^{-5} =
0.530734~M_\odot^2~{\rm pc}^{-5}\\ \\
\end{equation}
Thence we will quote $dJ/d\Omega$ in BUBU sr$^{-1}$ and $J$ in BUBU.  These units were chosen so that a cored isothermal profile for the Milky Way halo would have $dJ/d\Omega \sim 1$ in the direction of the Galactic Center.
\par
For a source whose size $R$ is small compared to its distance $D$,
we can replace 
$\ell$ in Eq.~(\ref{flux2})
by the source distance and use cartesian coordinates centered
at the source.  We write the volume element $dV = dx \,
dy \, dz$ where $z$ is along the line of sight and $x,y$ are transverse to the
line of sight. To study the angular dependence of the signal, we integrate in
$z$ only and introduce the angles $\theta_x = x/D$ and $\theta_y = y/D$.  In
terms of these, Eq.~(\ref{flux2}) gives
\begin{equation}
\frac{d\Phi}{dE d^2\theta} = 
\frac{1}{8\pi} \frac{dN_{\chi\chi
    \to \gamma}}{dE} \frac{\langle\sigma v\rangle}{m_\chi^2}
\frac{dJ}{d^2\theta}
\end{equation}
where
\begin{equation}
\frac{dJ}{d^2\theta} =
\int\rho^2 \, dz .
\label{flux3}
\end{equation}
\par
Integrating Eq.~(\ref{flux2}) over the small source ($R \ll D$) gives Eq.~(\ref{flux4})
with
\begin{equation}
  J = \frac{1}{D^2} \int\limits_{\rm source} \rho^2 \, dV.
\end{equation}
\subsection{Particle Model Examples}\label{particlemodels}
Particle physics enters the gamma-ray flux through the combination
\begin{equation}
\frac{dN_{\gamma}}{dE} \frac{ \langle \sigma v \rangle }{m_\chi^2}
\end{equation}
in Eq.~(\ref{flux2}).
We can estimate values for
$\frac{dN}{dE}$, $\langle \sigma v \rangle$, and the particle mass $m_\chi$ in
examples of WIMPs.  Once these values are given in a specific model, 
the resulting normalization required to fit the spectrum to the HESS flux
gives a value for $J$.  Varying the model parameters results in a band of $J$
values.
\par
We give here three examples of WIMPs: 
the lightest neutralino in minimal supergravity (mSUGRA), 
the lightest neutralino in a generic minimal supersymmetric standard model
(MSSM), 
and a Kaluza-Klein (KK) dark matter particle \cite{bergstrom05}.
\par
\begin{figure}[t]
  \begin{center}
    \includegraphics[height=25.pc]{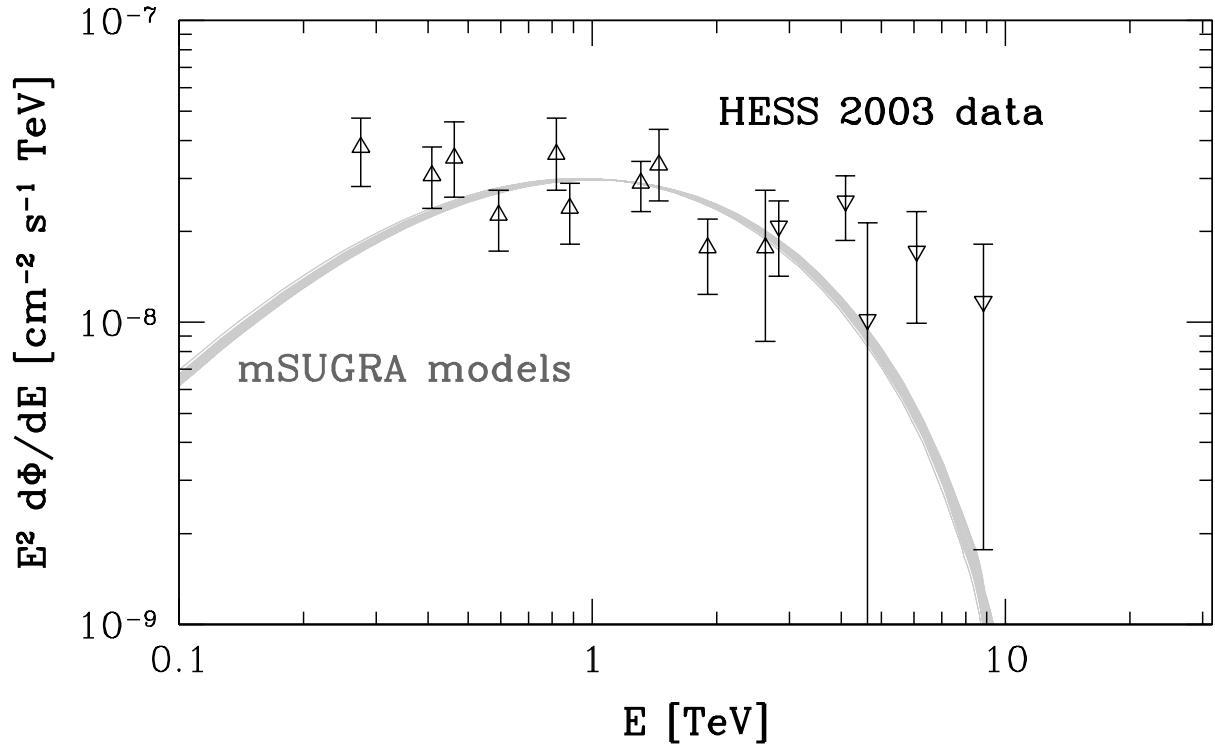}
  \end{center}
  \vspace{-0.5pc}
  \caption{\label{Fflux} Minimal supergravity models of the annihilation flux fit to
the 2004 HESS data. The grey band consists of several spectra generated with
mSUGRA neutralinos of mass $\sim 11$ TeV.}
\end{figure}

To explore mSUGRA models we used
the program
DarkSUSY \cite{gondolo04} to find model parameters consistent with
particle accelerator and direct search bounds.
The spectrum of gamma rays extends up to $\sim9$ TeV and any WIMP annihilation that would explain the observation would require a particle with a mass above $10$ TeV.  In mSUGRA excessive thermal relic densities are predicted for most neutralinos
with such a high mass.  However, changing the cosmological
model may alleviate this difficulty \cite{gelmini06,gelmini06_2}, so we proceed without
imposing the usual relic density constraint.  
We fit the normalization of the spectra to the HESS data.  The results
are shown in Fig.~(\ref{Fflux}).  The two physical processes included in this
spectrum are secondary pion decay and direct annihilation into photons.  
The spectral line due to direct photon production is not observable in the
spectrum after it has been convolved with 
the HESS energy resolution of $\sim15\%$.
Other
processes, especially the acceleration of charged secondaries, could be
reasonably expected to alter the spectrum \cite{2bergstrom05}. This could
provide other signatures of the 
annihilation which could be an important check on the DM
annihilation interpretation of the HESS flux.
We find that there is a family of mSUGRA models that produce a neutralino
with a mass of 10 to 11 TeV consistent with current constraints and have a
decent
agreement with the HESS spectrum, with a $\chi^2$ of $\sim$ 1.2.  These models 
require the $J$ parameter to be in the range 
[300,3000] BUBU to
explain
the flux observed by HESS.
\par
Lower values of $J$ may be obtained once the parameter space is relaxed beyond
mSUGRA.  The difficulty in finding mSUGRA models that fit the HESS data lies
in the excessive thermal relic densities predicted for neutralinos with the
required mass, $\sim 10$ TeV or higher.  Profumo \cite{profumo05} has suggested
that resonant annihilation of neutralinos through the A boson in the early
Universe, which can occur for $m_{\rm A} \simeq 2 m_\chi$, can lower the relic density for particles around 10 TeV.  In this case
the value of $J$ can be as low as $\sim1600$ BUBU (see his Fig.~(7b) obtained
in an anomaly mediated supersymmetry breaking model) or even
$\sim3$ BUBU (see his Fig.~8b, for a generic MSSM model).
\par
A third example of WIMPs that fit the HESS data are Kaluza-Klein particles.
Ref.~\cite{bergstrom05} finds that the spectrum for a model of KK DM
requires a $J$ value of $\sim130$ BUBU in order to be responsible for the flux
recorded by HESS.  
\subsection{DM density profile}\label{profilesection}
A dark matter density profile that would explain the TeV gamma-ray flux from
the Galactic Center with the particles discussed in the last section will need
to have a higher than expected density.
The HESS GC source does not extend beyond
$\sim0.1^\circ$ or $\sim10$ pc, covering a solid angle $\Omega \sim 10^{-5}$
sr.
The required $J$ value from section \ref{particlemodels} ranges from $3-3000$
BUBU or a $dJ/d\Omega \sim 10^5-10^8$ BUBU sr$^{-1}$ for $0.1^\circ$ or $\Theta
\sim 10^{-3}$ rad.
\par
The contribution from the extended DM halo along the line of sight to the GC
can be estimated from Eq.~(\ref{Jdef}) as $dJ/d\Omega \sim \rho^2 D$.  For a
canonical isothermal halo $\rho \sim 2~\rho_{local} \sim 0.6$ GeV cm$^{-3}$ and
$dJ/d\Omega \sim 4$ BUBU sr$^{-1}$ from the DM column through the extended DM
halo.  This is $5$ to $8$ orders of magnitude smaller than required, so higher
DM densities are needed to produce the gamma-ray flux by annihilation of our
candidate DM particles.
\par
A Navarro-Frenk-White profile (NFW) \cite{navarro97} is denser at the center.
For an NFW profile, the average value of $dJ/d\Omega$ within $0.1^\circ$ is,
from Eq.~(\ref{NFWJ}) below,
\begin{equation}
\frac{dJ}{d\Omega} \sim \frac{1}{\Omega} \int_\Omega \frac{dJ}{d\Omega} d\Omega = \frac{2 \pi \rho_s^2 r_s^2}{D \Theta} \sim 3
\times 10^3 ~{\rm BUBU~sr^{-1}}.
\end{equation}
  This is still $2$ to $5$ orders of magnitude too
small to explain the observed gamma ray flux for most of the dark matter
particles we consider.
We conclude that, if the HESS signal is due to DM annihilation, the extended
halo contributes no more than a few percent of the gamma-ray flux and a strong
enhancement in the density must exist within $10$ pc of the center of the
galaxy.
\par
The dark matter density profile within $10$ pc of the Galactic Center is not
known in detail and mechanisms for such density enhancements have been
proposed.  For example, such enhancement could be explained by extreme
clumping of the dark matter
\cite{silk93,kolb94,bergstrom99,aloisio02} which would have
implications on models of structure formation,
by steeper density profiles, $\rho \propto r^{-\alpha}$ with $\alpha \ge 3/2$,
have been suggested \cite{gurevich95} but are disfavored, or by a strong dark
matter
concentration at the Galactic Center (a spike
\cite{gondolo99,ullio01,merritt02,gnedin04,bertone05,2bertone05}).
To include the latter two possibilities, we split the dark matter profile into
an inner and an outer part at a transition radius R$_I$.
\par
As an example of the outer profile we use the NFW
profile
\begin{equation}
\label{NFWProfile}
\rho_{\rm NFW} = \frac{\rho_s}{\frac{r}{r_s}\left[ 1+\left(\frac{r}{r_s}\right)^2\right]}.
\end{equation}
$r_s$ is a scale radius and $\rho_s$ is twice the density at $r_s$.  We will
take typical values \cite{bergstrom98} of  $r_s =
25$ kpc and $\rho_s = \rho_0 (D/r_s) [1+(D/r_s)^2]$ with a local density
$\rho_0 =
0.3$ GeV cm$^{-3}$. We take the distance to the Galactic Center to be $D=8$
kpc \cite{reid93}.
For this profile we compute
\begin{equation}
\frac{dJ_{\rm NFW}}{d\Omega} = \rho_s^2 r_s \left\{ \frac{\pi - \theta}{y} -
  \frac{3+2y^2}{2(1+y^2)}\left[\arctan\left(\frac{z}{\sqrt{1+y^2}}\right)+\frac{\pi}{2}\right] -
  \frac{z}{2(1+x^2)(1+y^2)}\right\}
\label{NFWJ}
\end{equation}
where $\theta$ is the angle between the line of sight and the GC, $x = D/r_s$,
$y=x\sin\theta$, and $z=x\cos\theta$. 
\par
Notice that Eq.~(\ref{NFWJ}) diverges in the direction of the GC ($\theta = 0,
y=0$) as $\pi \rho_s^2 r_s^2 / D \theta$.
To remove the inner part of the NFW profile, we add an inner cutoff at $R_{\rm
  I}$ by replacing the term
$(\pi - \theta)/y$
in Eq.~(\ref{NFWJ}) with 
\begin{equation}
{\cal F}(y,z_c) + {\cal F}(y,b_c) -
  \frac{z_c}{(1+y^2)(1+x_c^2)} - \frac{3+2y^2}{(1+y^2)^{\frac{3}{2}}}
  \arctan \left( \frac{z_c}{\sqrt{1+y^2}} \right) 
\end{equation}
where $x_c=R_{\rm I}/r_s$, $z_c = \sqrt{x_c^2 - y^2}$,
$b_c = (z z_c+y^2)/(z-z_c)$, and
\begin{equation}
{\cal F}(y,a) = \frac{1}{\sqrt{a^2+y^2}}
\,{}_2F_{1}\!\left(\frac{1}{2},\frac{1}{2};\frac{3}{2};\frac{y^2}{y^2+a^2}\right) =
\frac{1}{y} \arctan\!\left(\frac{y}{a}\right)
\end{equation}
The form with the hypergeometric function is used to avoid division by zero at
$y=0$ ($\theta = 0$).
\par
In the inner ($r \lesssim 10$ pc) part of the density profile we use a simple
functional form
to model a central density enhancement.
We assume that a DM mass $M_{\rm I}$ is contained within a sphere of radius
$R_{\rm I}$, and that its density profile is spherically symmetric and
decreases with a power $\alpha$ of the radius. 
The inner profile we use is
\begin{equation}
\label{DMball}
\rho_{\rm I}(r) = \begin{cases}
\displaystyle
\frac{3-\alpha}{4\pi}\, \frac{M_{\rm I}}{R_{\rm I}^3}
\left(\frac{r}{R_{\rm I}}\right)^{-\alpha}, & R_c \le r \le R_{\rm I}, \\
0, & {\rm otherwise}.
\end{cases}
\end{equation}
This inner profile could be a steep profile, or a spike around the central
black hole.
For this inner profile, we find 
\begin{equation}
J_{\rm I} = \frac{(3-\alpha)^2}{4\pi} \frac{M_{\rm I}^2}{R_{\rm I}^3 D^2} \frac{1}{3-2\alpha}
\left[ 1 - \left(\frac{R_c}{R_{\rm I}}\right)^{3-2\alpha} \right].
\end{equation}
Here R$_c$ is an inner cutoff radius discussed in the next paragraph. 
For the angular profile we compute
\begin{equation} \label{angularprofile}
j(\theta) \equiv \frac{1}{J}\frac{dJ(\theta)}{d^2\theta} = \begin{cases}
\displaystyle
\frac{1}{2\pi}\frac{3-2\alpha}{\theta_{\rm I}^{3-2\alpha}-\theta_c^{3-2\alpha}} 
\Bigg[
\frac{\theta_{\rm I}^{1-2\alpha}}{1-2\alpha}
\,F\!\left(\alpha,\frac{\theta}{\theta_{\rm I}}\right)
-
\frac{\theta_{c}^{1-2\alpha}}{1-2\alpha}
\,F\!\left(\alpha,\frac{\theta}{\theta_{c}}\right)
\Bigg],
& \theta < \theta_{c},
\\
\displaystyle
\frac{1}{2\pi}\frac{3-2\alpha}{\theta_{\rm I}^{3-2\alpha}-\theta_c^{3-2\alpha}} 
\Bigg[
\frac{\theta_{\rm I}^{1-2\alpha}}{1-2\alpha}
\,F\!\left(\alpha,\frac{\theta}{\theta_{\rm I}}\right)
-
\frac{\theta^{1-2\alpha}}{1-2\alpha}
\frac{\sqrt{\pi} \ \Gamma(\alpha+\frac{1}{2})}{\Gamma(\alpha)}
\Bigg],
& \theta_{c} < \theta < \theta_{\rm I},
\end{cases}
\end{equation}
and zero for $\theta > \theta_{\rm I}$.
Here we defined $\theta_{\rm I} = R_{\rm I} / D$, $\theta_c = R_{c} / D$, and
\begin{equation}
F(\alpha,x) = \sqrt{1-x^2}
 \ _2F_1(\alpha,1;\alpha+\frac{1}{2};x^2)
\end{equation}
where $_2F_1$ is the hypergeometric function.  Notice that for $\alpha = 3/2$
the factor in front of the square bracket is $[2 \pi \ln(\theta_{\rm I} / \theta_c]^{-1}$ and
that for $\alpha = 1/2$ the square bracket is $\ln(\theta_{\rm I} /
\theta_{c})$.
\par
An inner cutoff at
$R_c$ is introduced to avoid the divergence that occurs in $dJ_{\rm
  I}/d\Omega$ when $\alpha \ge 3/2$.   
This inner cutoff is left as a free parameter, because
this part of the halo is even more unknown than the rest.
Physically an inner
cutoff would naturally be present.  Either the capture radius of the black
hole, or some effective radius at which, e.g., the DM density is depleted by
annihilation during the history of the Milky Way.   
In the latter case the maximum sustainable density is usually taken as 
\begin{equation}\label{maxdensity}
\rho~\le~\rho_{max}~\simeq~\frac{m}{\langle \sigma v \rangle t} 
\end{equation}
with the time $t$ taken as the age of the Milky Way.
In the case we are considering, we have a measurement of the flux and of the
particle
mass from the extent of the spectrum. For example, integrating
Eq.~(\ref{flux4}) in energy above the threshold and inserting
Eq.~(\ref{DMball}) with $\alpha = 0$, Eq.~(\ref{maxdensity})
implies
\begin{equation} 
M_{\rm I}~\ge~M_c~\equiv~\frac{8 \pi D^2 m t \Phi}{N_{\gamma}},
\end{equation}
where $\Phi$ is the total photon flux above threshold and $N_{\gamma}$ is the
number of photons produced above threshold in each annihilation.
Thus the maximum density $\rho_{max}$ corresponds to a lower limit on the
mass
of an inner feature of the halo that could explain the gamma-ray
observation.
If the mass is too small, then the cross-section and density required to
mantain the same flux are so large that the feature would have
annihilated by now.
This can be generalized to all $\alpha$ values by finding the $R_c$
for which $\rho(R_c) \le \rho_{max}$.  
They are any $R_c$ greater than the solution for $R_q$ in the equation
\begin{equation}\label{rhomax}
R = R_q \left( 1 + \frac{M_c^2}{D^2 J} \frac{3 - 2\alpha}{4 \pi R_q^3}
\right)^\frac{1}{3-2\alpha}. 
\end{equation}
\par
Another scale in this problem is the capture radius of a
3 million solar mass black hole, expected to be in the center of all of these
profiles.  We find that the capture radius, $\sim 10^{-7}$ pc, is greater than
all
$R_q$.
\par
Thus, as a physically motivated number, we take the range of cutoff radii to
be
\begin{equation}
10^{-7}~{\rm pc} \le R_c \le R_{\rm I}.
\end{equation}
\subsection{Limits from the HESS angular profile}\label{HESSangles}
The angular distribution of photons in the HESS detector carries information
on the source profile.  Here we investigate the constraint on the source
profile due to these data.
\par
The HESS analysis \cite{aha04} assumes a gaussian source profile, and gives a limit on
the source angular size equal to $\leq3'$. To determine the limit on our power-law sphere
in Eq.~(\ref{DMball}), we compare the emission profile,
Eq.~(\ref{angularprofile}),with
the angular distribution of detected photons.
Fig.~(2) in \cite{aha04} gives the
photon counts $C_i$ and their errors $\delta_i$ in each $\theta_{i}^{2}$
bin. Here
$\theta_{i}$ is the angle between the photon direction and the center of the
excess.  The center of the excess agrees to the position of the GC to well
within the systematic errors in the pointing of the HESS array.
The intrinsic angular profile 
\begin{equation}
j(\theta) = \frac{1}{J} \left[ \frac{dJ_{\rm I}}{d\Omega} + \frac{dJ_{\rm o}}{d\Omega} \right]
\end{equation}
is convolved with the
point spread function (psf) of HESS as given in \cite{horns05}:
\begin{equation}
f_{psf}(\theta) =
f_0 \left[ e^{-\frac{\theta^2}{2 \sigma_1^2}} +
\frac{1}{8.7}e^{-\frac{\theta^2}{2 \sigma_2^2}} \right].
\end{equation}
Here $f_0$ was chosen so the psf has unit area and the widths of the gaussians
are $\sigma_1$ = 0.052$^\circ$ and $\sigma_2$ = 0.136$^\circ$.
We rewrite this as a linear combination of two normalized gaussians,
\begin{equation}
f_{psf}(\theta) = \frac{c_1}{2 \pi \sigma_1^2} e^{-\frac{\theta^2}{2 \sigma_1^2}}
+ \frac{c_2}{2 \pi \sigma_2^2} e^{-\frac{\theta^2}{2 \sigma_2^2}}
\end{equation}
with $c_1 = 8.7 \sigma_{1}^{2} / (8.7 \sigma_{1}^2 + \sigma_{2}^2)$ and $c_2
= \sigma_{2}^{2} / (8.7 \sigma_{1}^2 + \sigma_{2}^2)$. 
The source profile convolved with a normalized gaussian
is
\begin{equation}
\overline{j}(\theta,\sigma) = \sigma^{-2} e^{-\frac{1}{2}(\frac{\theta}{\sigma})^2}
\int_{0}^{\theta_{DM}} \theta^\prime
e^{-\frac{1}{2}(\frac{\theta^\prime}{\sigma})^2}
I_0\Big(\frac{\theta \theta^\prime}{\sigma^2}\Big)
j(\theta^\prime)  d\theta^\prime
\end{equation}
and $j$ convolved with the entire psf is
\begin{equation}
j_{psf}(\theta) = c_1 \overline{j}(\theta,\sigma_1) + c_2 \overline{j}(\theta,\sigma_2).
\end{equation}
 The photon counts $C(\theta)$ as a function of angle $\theta$ from the GC are
 proportional to the convolution of $dJ/d\Omega$ with the psf
\begin{equation}
C(\theta) = A \left( \frac{dJ}{d\Omega} \right)_{\rm psf}.
\end{equation}
The proportionality constant is given by
\begin{equation}
A = {\cal E} \frac{\langle \sigma v
  \rangle}{m_\chi^2} \frac{N_\gamma}{8 \pi} \pi \Delta \theta^2,
\end{equation}
where ${\cal E}$ is the exposure, $N_{\gamma}$ is the total number of
photons
\begin{figure}
  \begin{center}
    \includegraphics[height=25.pc]{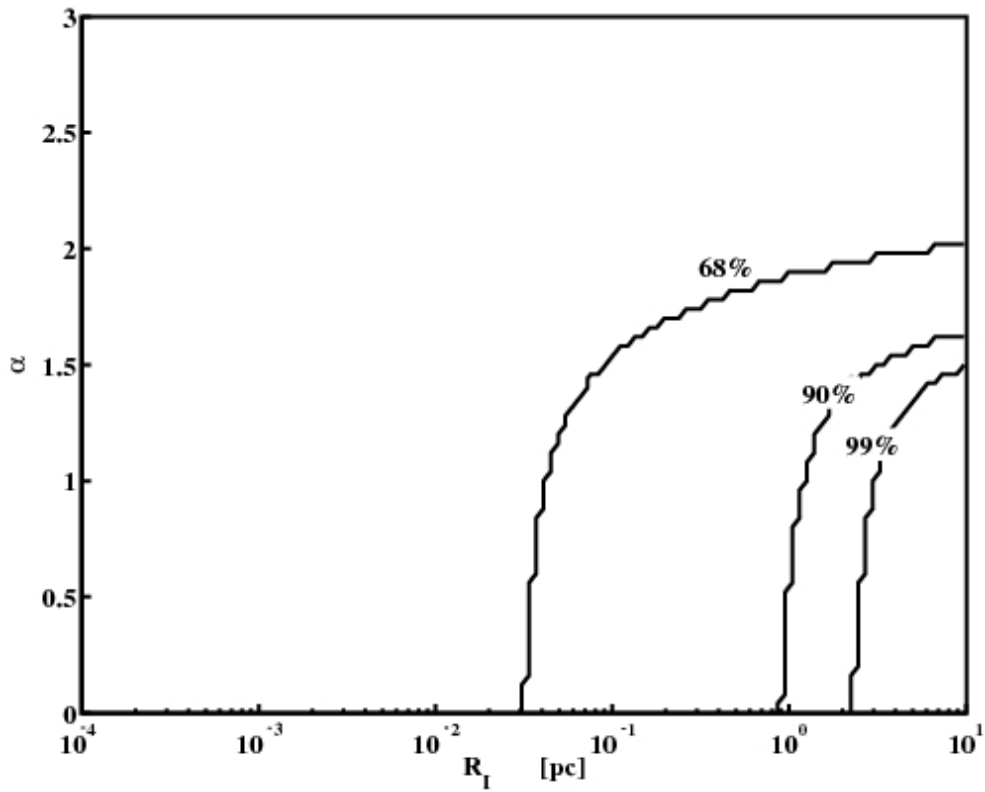}
\end{center}
  \caption{\label{rvsalphaHESS} The constraint from the HESS angular
    size.  The profiles are generally required to be steep, $\alpha>1.5$,
    or small, $R_{\rm I}<1$ pc.}
\end{figure}
above the experimental threshold emitted per annihilation, and $\Delta
\theta^2$ is the aperture of the observation.  We have
estimated the HESS exposure as the ratio of the total counts assuming a point
source and the integral flux of $1.82 \times 10^{-6} $m$^{-2} $s$^{-1}$ above
threshold, both as reported by HESS in \cite{aha04}. We
estimate an  
exposure ${\cal E}
\sim 3 \times 10^{13}$ cm$^2$ s.
Furthermore,
\begin{equation}
\left( \frac{dJ}{d\Omega} \right)_{\rm psf} = J \, j_{\rm psf}(\theta) .
\end{equation}
The best fit
for the normalization factor $A$ is 
\begin{equation}
A = \sum_i \frac{C_i 
\, J \, j_{psf}(\sqrt{\theta^{2}_{i}})}{\delta_i},
\end{equation}
with the $\chi^2$ given by
\begin{equation}
\chi^2 = \sum_i \Bigg(\frac{C_i - A J j_{psf}(\sqrt{\theta^{2}_{i}})}{\delta_i}\Bigg)^2.
\end{equation}
\par
Here to find our intervals we perform a bayesian analysis.  We take the 
likelihood as proportional to $e^{-\chi^2/2}$ and
define our
confidence intervals in $(\alpha,R_{\rm I})$ as the corresponding quantiles
of the posterior distribution.  We take the
prior
distribution as flat in $\log R_{\rm I}$ and $\alpha$, and zero outside
the range shown in Fig.~(\ref{rvsalphaHESS}).
\par
The intermediate results of this piece of the analysis are shown in
Fig.~(\ref{rvsalphaHESS}). 
The $68\%$, $90\%$, and $99\%$ regions in the ${R_{\rm I}, \alpha}$ are shown.  At the $90\%$ 
confidence level the HESS data confines the source diameter to $\lesssim 1$ pc for
a uniform sphere.  For power law
density profiles with index $\alpha \gtrsim 1.5$ the constraint on the source
size starts to weaken considerably; these profiles could be modeled as
a smaller source with a harder power law index.
\begin{figure}
  \begin{center}
    \includegraphics[height=25.pc]{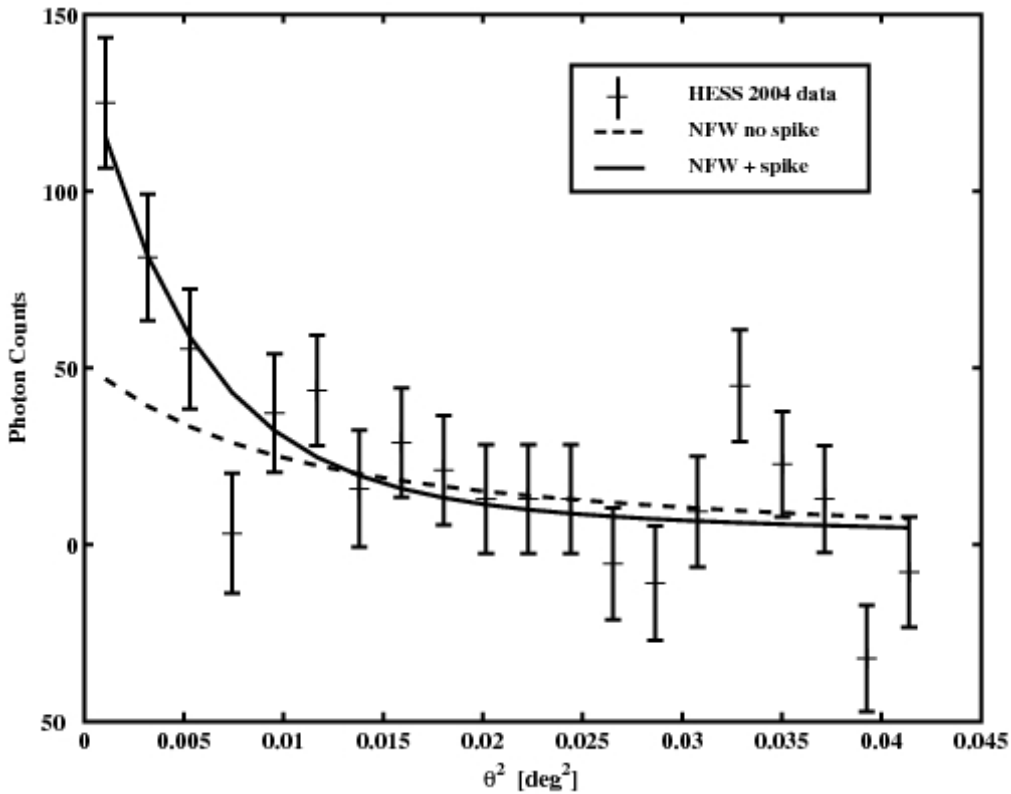}
\end{center}
  \caption{\label{NFWFit} 
  An NFW+spike fit to the 2004 HESS gamma-ray data on the Galactic Center.  For
  this fit we used a NFW profile outside $1$
  pc and a $\rho \sim r^{-1.9}$ profile for $r < 1$ pc.  An inner cutoff of
  $10^{-5}$ pc was used to approximate the inner core.  This is a typical
  profile
  of a spike after dynamical heating
  and annihilation effects are included.
  }
\end{figure}
Finally, we include a fit of two profiles in Fig.~(\ref{NFWFit}).  The first,
more shallow profile is an NFW profile alone.  Evidently the flux rules out an
NFW only DM profile, and so does the angular profile of the gamma-ray source.
The steeper profile is a fit of an NFW profile with a spike.  A spike can
simulate a point source effectively, and the extended NFW piece makes the
profile fit slightly better than a point source.
\par
As an illustration of the kind of angular
gamma-ray profiles compatible with the HESS data in models with a DM spike we
show such a profile for an NFW model.  We fix the halo NFW model with the parameters give in section \ref{profilesection}.  A pure NFW model does not fit the
data well at small angles from the GC.  However, motivated by
\cite{gondolo99}, we add a spike with a radius of $1$ pc and a slope of
$\alpha = 1.9$, which are
typical values after inclusion of the effects of stars and annihilation
\cite{bertone05}.
The $\chi^2$ per degree of freedom for the fit
is $1.3$ which is the same ($\Delta \chi^2 \sim 10^{-4}$) as in the fit of
a point-like source. Since the density of the dark matter is fixed for this
profile is fixed to the local density and the particle mass is bounded by the
spectrum, the normalization of the fit to the
flux gives the value of
the cross-section, $N_{\gamma} \langle \sigma v \rangle = 3.3 \times
10^{-28}$ cm$^3$ s$^{-1}$. This is a reasonable value for the WIMP
annihilation cross section in particle physics models.
\subsection{Limits from stellar dynamics mass measurements} \label{massmeasure}
Measurements of the amount of mass $M(< r)$ contained within a distance $r$
from the GC are continuously improved as more precise data are collected. Here
we use the compilation of enclosed mass measurements in \cite{ghez03}. From
their paper, we extract a table of the mass $M_i$ contained within radius
$r_i$, together with its quoted error $\sigma_i$.

To these data we fit a mass profile with three components: the central black hole, the central stellar cluster, and the dark matter sphere described above.
\begin{equation}
M(< r)  =  M_{\rm BH}  + M_{*}(< r) + M_{\rm I}(< r).
\end{equation} 
$M_{\rm BH}$ is the mass of the central black hole Sagittarius A$^*$. 
For the central stellar cluster we use the empirical mass profile $M_{*}(< r)$
obtained from data in~\cite{genz03},
\begin{equation}
M_{*}(< r) = \begin{cases}
\displaystyle M_{*} \left(\frac{r}{R_*}\right)^{1.6},    &   r \le R_*,\\
\displaystyle M_{*} \left(\frac{r}{R_*}\right)^{1.0},    &   r > R_*,
\end{cases}
\end{equation}
with $M_{*} = 0.88 \times 10^6 M_{\odot}$ and $R_*$ = 0.3878 pc. We model the
dark matter with the density profile described in Eq~(\ref{DMball}), which corresponds to a DM mass profile
\begin{equation}
M_{\rm I}(< r) = 
\begin{cases}
\displaystyle M_{\rm I} \left(\frac{r}{R_{\rm I}}\right)^{3-\alpha},    &   r \le R_{\rm I},\\
\displaystyle M_{\rm I},    &   r > R_{\rm I} .
\end{cases}
\label{DMBall}
\end{equation}
We use the likelihood function to find constraints on the DM density
profile, using a bayesian analysis similar to section \ref{HESSangles}. 
Assuming
the errors quoted in~\cite{ghez03} are gaussian, the likelihood ${\cal L}$ is
given by
\begin{equation}
2 \ln {\cal L} = \sum_i \frac{(M_{i} - M(< r_i))^2}{\sigma_i^2} = \sum_i
\frac{(M_{i} - M_{\rm BH}  - M_{*}(< r_i) - M_{\rm I} f_i)^2}{\sigma_i^2}
\label{L}
\end{equation}
with 
\begin{equation}
f_i = 
\begin{cases}
\displaystyle \left(\frac{r_i}{R_{\rm I}}\right)^{3-\alpha},    &   r_i \le R_{\rm I},\\
\displaystyle 1,  &   r_i > R_{\rm I} .
\end{cases}
\end{equation}
In order to obtain a constraint on the parameters
$M_{\rm I}$ and $R_{\rm I}$ at a fixed value of $\alpha$, we first marginalize over
$M_{BH}$.  Since $\ln {\cal L}$ is quadratic in $M_{BH}$, we need only replace $M_{BH}$
in Eq.~(\ref{L}) with the value $M_{BH}^0$ obtained by maximizing the
likelihood. This is given by
\begin{equation}
M_{BH}^0 = \frac{x_3x_4-x_2x_5}{x_3^2-x_1 x_5},
\end{equation}
with
\begin{equation}
x_1 =\sum \frac{1}{\sigma_i^2} ,  
\quad
x_2 = \sum \frac{M_i - M_*(< r_i)}{\sigma_i^2} , 
\quad
x_3 = \sum \frac{f_i}{\sigma_i^2}, 
\quad
x_4 = \sum \frac{f_i[ M_i-M_*(< r_i)]}{\sigma_i^2},
\quad
x_5 = \sum \frac{f_i^2}{\sigma_i^2}. 
\end{equation}
As our prior, $R_{\rm I}$ is restricted to the range 
[0.0004,10] pc and distributed logarithmically,
and $\alpha$ is kept at a few fixed values (0,1,2).
By integrating our posterior probability distribution 
we derive a 1 sigma upper limit and a 90$\%$ bayesian interval
in the $M_{\rm I}$, $R_{\rm I}$ parameter space.
For $R_{\rm I}$ smaller than the innermost data point (0.0004 pc), 
there is a degeneracy
between $M_{BH}$ and $M_{\rm I}$.  
We break this degeneracy by imposing $M_{\rm I}$
less than the upper bound on the black hole mass reported in \cite{ghez03} 
($3.6 \pm 0.4 \times 10^6 M_\odot$).
This is equivalent to assuming all the mass within the innermost
orbit could be DM.
\section{Results}
\par
From the particle examples in Section \ref{particlemodels} we find that a
range of $J = [300,3000]$ is
needed to explain the flux of gamma-rays from the GC as DM annihilation
products.  With resonant annihilations, $J$ can be as low as $\sim 1$.
Furthermore we conclude that the DM annihilation line will be unobservable
with an energy resolution of $10\%$.  We find that the spectrum of gamma rays
from the GC is compatible with the decay of pions produced in the
annhilation.  More complex models of the radiation, such as \cite{2bergstrom05}
where bremsstrahlung of $W$ products has an appreciable effect, the
spectrum may be similar to a power law and other spectral features, such as a
hardening of the spectrum near the WIMP mass, may be observable.  
\par
The
requirement that the central feature of dark matter not annihilate in the
lifetime of the Universe gives a lower limit on the mass.  For example, a
central feature with $\alpha = 0$ and an upper limit on the density of $\rho
= 10^{15}$ M$_{\odot}$ pc$^{-3}$ and the requirement that $J = 1000$ BUBU gives
a lower limit of $\sim 3 \times 10^{-4}$ M$_{\odot}$.  For a limit density of
$\rho = 10^{12}$ M$_{\odot}$ pc$^{-3}$ the mass of annihilating dark matter
must
be greater than $\sim 1$ M$_{\odot}$ to be stable for $10^{10}$ years.  
These limits are below the
lower edge of our results plots.
\par
The stellar dynamics limit extended mass distributions to $\sim10\%$ of the
black hole mass for $R_{\rm I} = (10^{-3},1)$.  The angular size bounds are
complementary excluding 
regions above a radius that depends on the assumed $\alpha$ for the
distribution, as seen in Fig.~(\ref{rvsalphaHESS}).
\par
\begin{figure}
  \includegraphics[height = 25.pc]{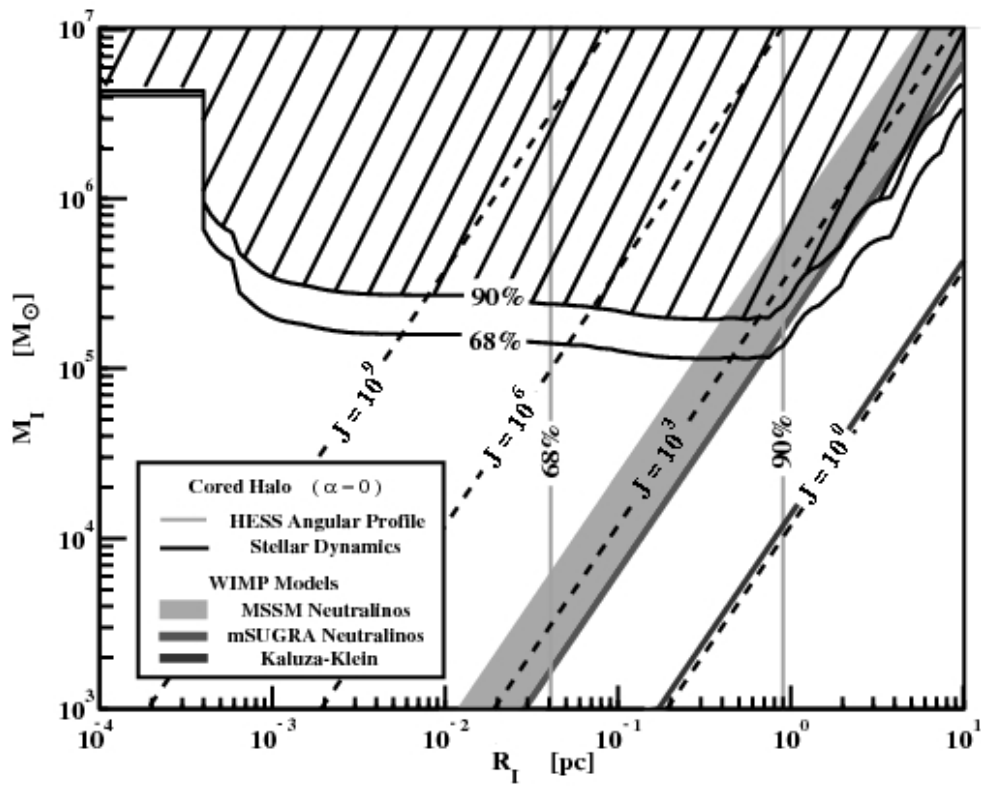}
  \caption{\label{F4} Stellar orbit constraints on the mass and size of a dark matter concentration at the Galactic center. The black lines are the $1$ sigma upper limit and $90\%$
  confidence interval for the mass
  of dark matter spheres with power law density profile index $\alpha = 0$.  The dashed
  lines show the mass corresponding to the $J$ 
  values indicated (in BUBU).  The $1$ sigma and $90\%$ bounds from the angular
  profile are
  shown as the vertical gray lines.  The grey bands show the typical values of
  $J$
  required to produce the HESS flux in WIMP models: 
  KK (dark grey), mSUGRA (medium grey), and MSSM (light grey) }
\end{figure}
\begin{figure}
  \includegraphics[height = 25.pc]{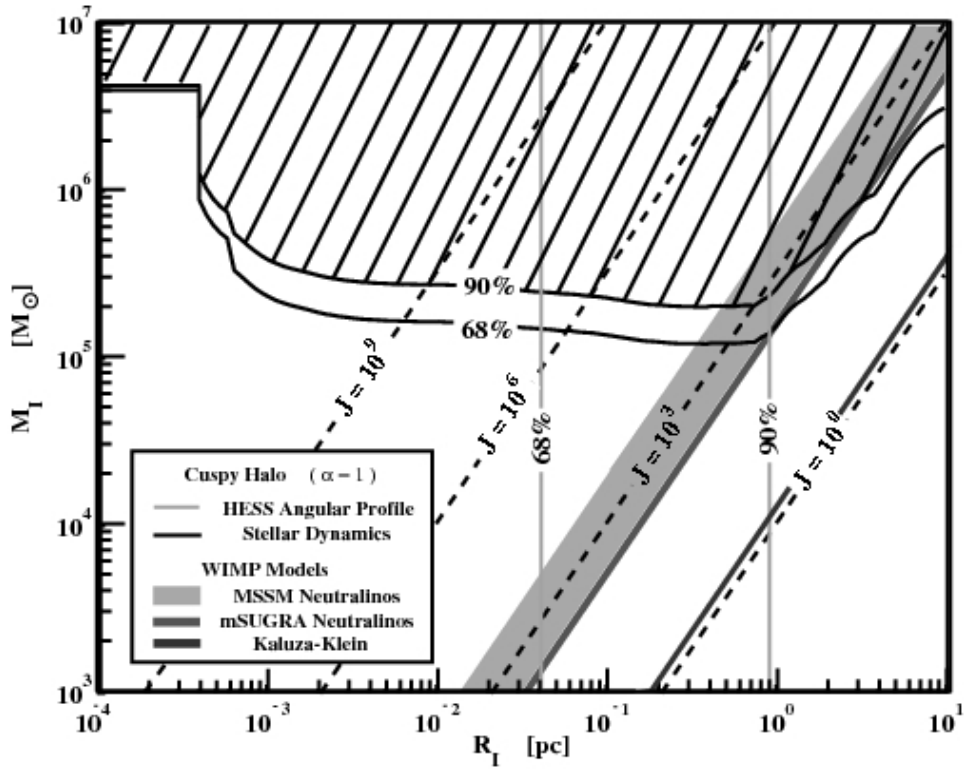}
  \caption{\label{F5} Same as Fig.~(\ref{F4}) but for dark matter spheres with power law
  density profiles with $\alpha = 1$. }
\end{figure}
\begin{figure}
  \includegraphics[height = 25.pc]{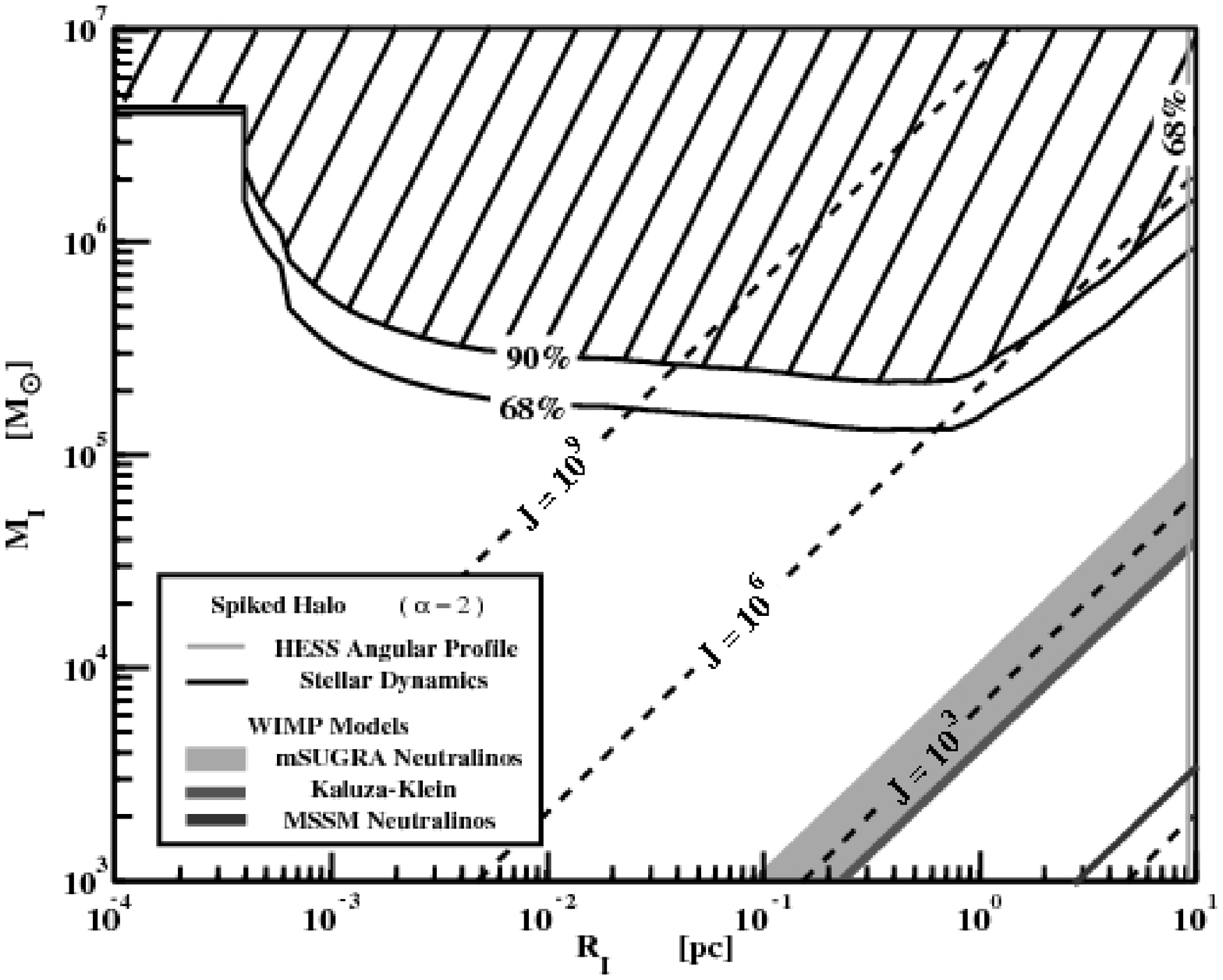}
  \includegraphics[height = 25.pc]{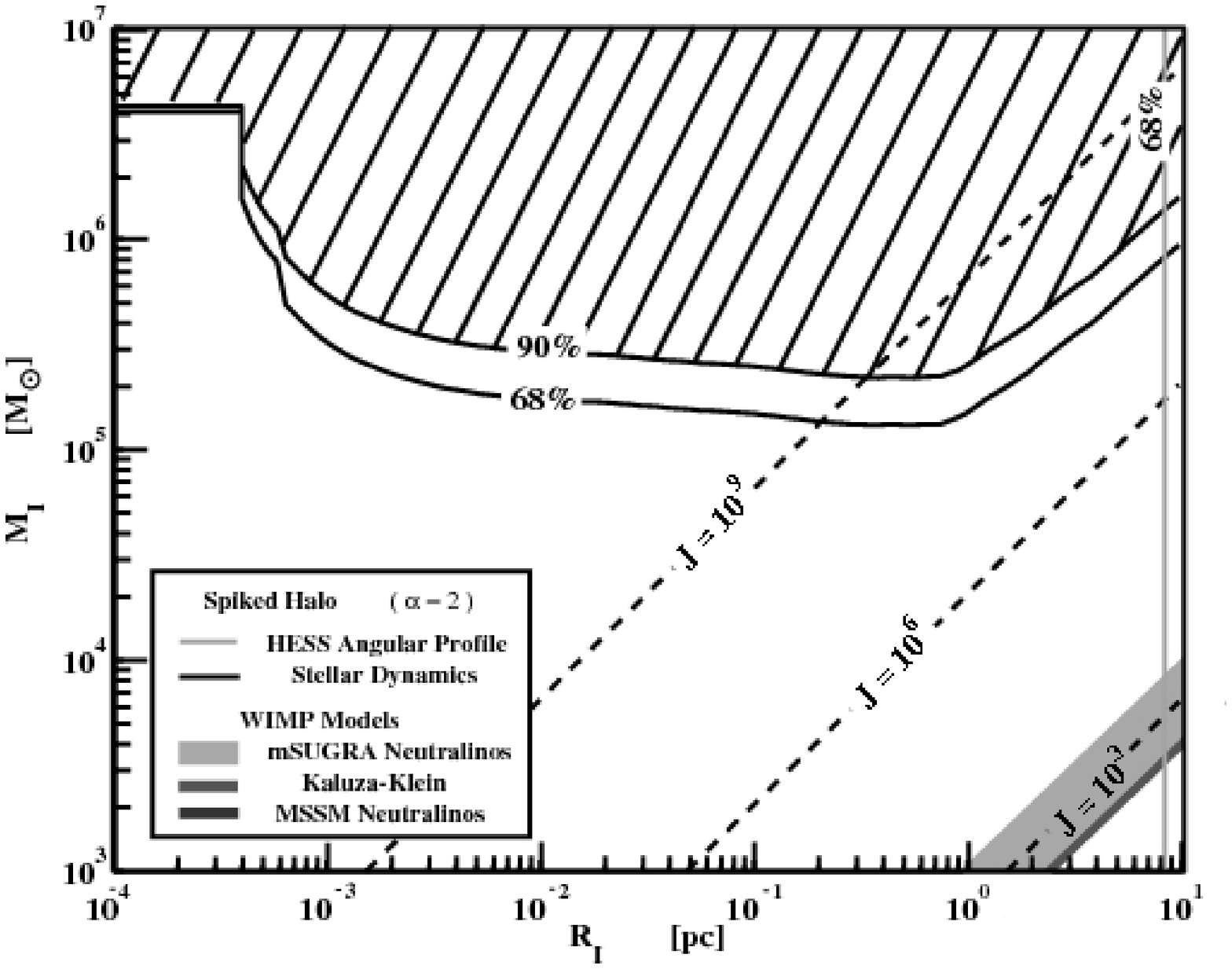}
  \caption{\label{F6} Same as Fig.~(\ref{F4}) but for inner DM profiles
  with $\alpha = 2$. For this steep density profile, which
  ``looks'' like a smaller sphere, the size constraint on
  the DM feature is not as strong, constraining values of $J\sim10^6$ BUBU or
  higher.  To avoid a divergence in $J$ and because these profiles are
  expected to flatten at some inner radius, these profiles were computed with
  a minimum cutoff radius $R_c = 10^{-4}$ pc (top) and $R_c = 10^{-6}$ pc
  (bottom).}
\end{figure}
\begin{figure}
  \begin{center}
    \includegraphics[height=25.pc]{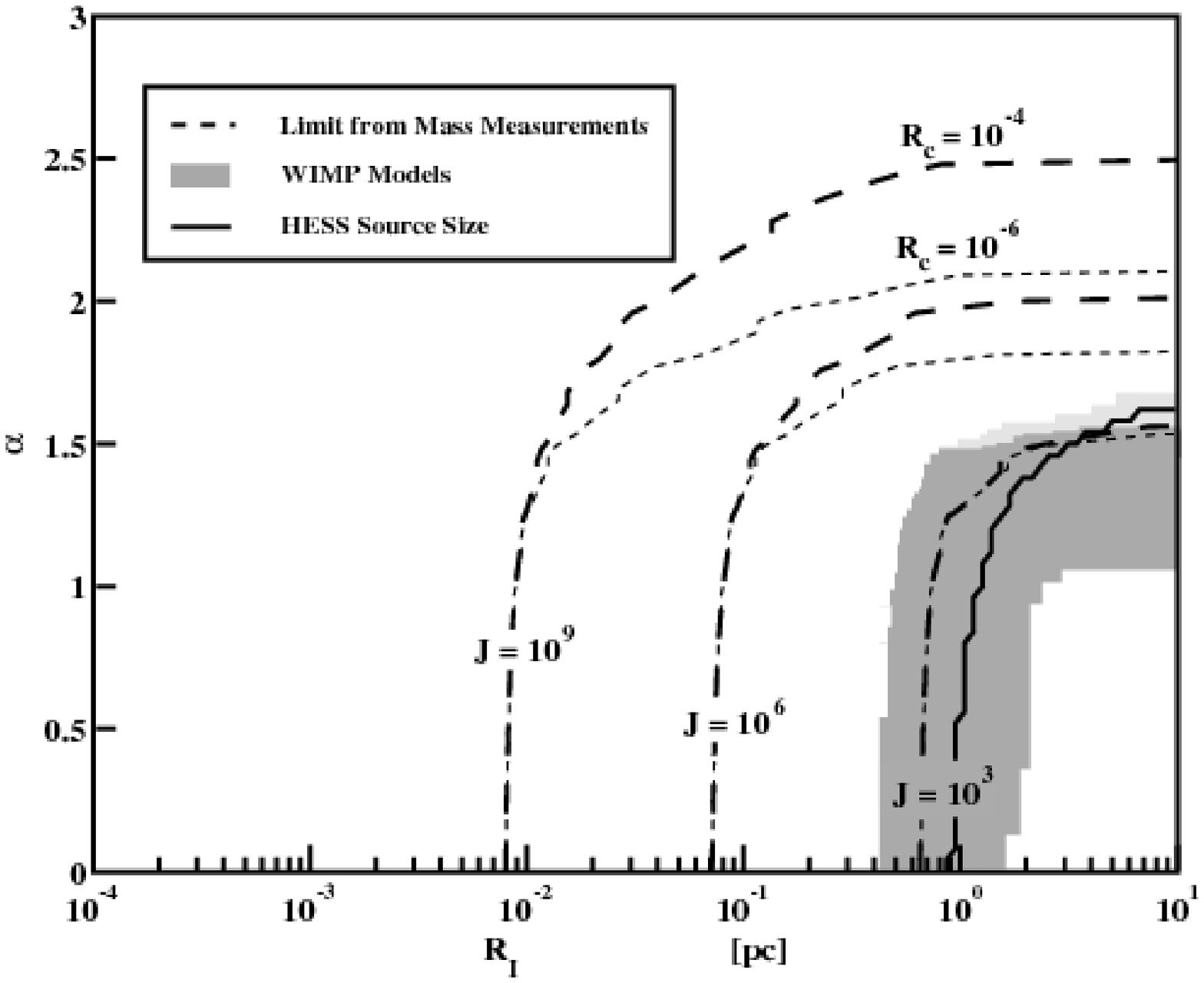}
\end{center}
  \caption{\label{rvsalpha} 
    Constraints on the $J$ parameter due to stellar orbit data
    \cite{ghez03,genz03} and the HESS source profile.
    The dashed lines show the crossing of specific $J$ values with the
    $90\%$ confidence level from stellar orbit
    measurements as shown in Figs.~(\ref{F4},\ref{F5}, and \ref{F6}) for
    specific values of $\alpha$.  
    The $90\%$ constraint on the source profile from the HESS data
    alone is shown as the solid line.  The grey band is the expected range
    of $J$ for mSUGRA neutralinos that produce the GC TeV flux.}
\end{figure}

The results of the analysis are compiled in Figs.~(\ref{F4}, \ref{F5},
\ref{F6}, and \ref{rvsalpha}).
In Figs.~(\ref{F4}, \ref{F5}, \ref{F6}) we plot both the stellar dynamics
bound and the angular size bound in the $M_{\rm I}-R_{\rm I}$ plane for
three values of $\alpha$.
The expected range from the particle physics are shown as shaded regions.
These
regions correspond to either a value of $J$ that could produce the observed
flux, or equivalently to a value for
$N\langle \sigma v \rangle$.
\par
Comparing these models to our stellar dynamics bounds,
we see that for $\alpha=0$ (Fig.~\ref{F4}) the source size is restricted to
$\lesssim 20$ pc
in mSUGRA and KK models.  For larger cross-sections with resonant
annihilation, the source size is unbounded by the stellar dynamics.
The constraint from the HESS source profile limits the source size to
$\lesssim 1$ pc (vertical line),  so it is similar to the stellar dynamics
constraint in mSUGRA and KK models, but is stronger for resonant-annihilation
models.  However, for some of the mSUGRA models we considered the stellar
dynamics constrains were stronger restricting the source size to $\lesssim
0.3$ pc.
\par
For profiles with shallow cusps ($\alpha=1$; Fig.~\ref{F5}), the source size
constraint on WIMP models from stellar dynamics
is similar to the $\alpha=0$ case. No bounds for resonant-annihilation models,
but
still $\lesssim 20$ pc for mSUGRA and KK models.  The HESS constraint from the
angular
size of the gamma-ray excess is still $\sim 1$ pc and so conclusions similar to
those with $\alpha=0$ apply in this case.
\par
For profiles with steep cusps ($\alpha=2$; Fig.~\ref{F6}) stellar dynamics
bounds out to $10$ pc do
not provide a constraint on the WIMP models we examined.  The constraint from
the HESS angular profile is also much weaker here.  
We show two plots here to illustrate the effect of the cutoff radius which
only comes into play for these steep profiles.  We show two cutoff radii of
$10^{-4}$ and $10^{-6}$ pc.
\par
The constraints from stellar orbits and the HESS angular distribution are summarized for comparison in figure \ref{rvsalpha}.  The solid line represents the $90\%$
confidence region based on the HESS data alone.  The dotted lines show the
constraint coming from stellar dynamics.  
Various values of $J$ are plotted so that these constraints can be compared
to particle physics models.
The values of $J$ required by the mSUGRA neutralinos, MSSM neutralinos, and Kaluza-Klein particles we examined are plotted as
the medium grey band.  Both the stellar dynamics and the gamma-ray angular
profile point to a DM source that is either small or steep.
\par
\par
\section{Conclusions}
To summarize, a very high energy gamma-ray flux from the center of the Milky 
Way was significantly detected by the HESS collaboration
during 2003-2004.
A possible explanation of the very high energy radiation from the Galactic Center is WIMP annihilation.  The intensity of the annihilation flux
is a function of the density profile of dark matter in the galactic
center.  The angular distribution of detected gamma rays limits the size of the emission region. Data on 
on the proper motions of stars and star counts around the galactic
center constrain the 
size and the mass of the dark matter at the GC. We have shown that the density needed to produce
the observed flux from WIMP annihilation is consistent with observational constraints 
on the mass profile of the GC.
For the stellar orbit data and the star counts, we used the infrared data in \cite{ghez03} and \cite{genz03}.
We found that these astronomical
constraints on the source profile are comparable to and slightly stronger than
the constraint from the angular distribution of photons measured by HESS. 
\par
\par
There are several ways in which WIMP annihilation as the origin of the HESS
flux could be confirmed or made implausible. As is clear from
fig.~\ref{rvsalpha}, a slight improvement in either the gamma-ray angular
resolution or the constraints from stellar orbits may reveal the presence of
an extended dark matter annihilation region at the Galactic Center. An
extended emission out to large angles would be a possible indication of WIMP
annihilation. An extended gamma-ray excess with the same spectrum and position
of the GC flux has recently been
reported by HESS \cite{aha06}.
A spectral cutoff at energies higher than the
particle mass is another requirement of the DM
hypothesis.  The cutoff may be preceded by a gamma-ray line at the particle
mass, but this
 spectral line does not appear to be
observable with atmospheric Cherenkov telescopes in the particle models we examined due to the insufficient energy resolution.
Absence of variability is another feature of WIMP annihilation, thus variability of the source
would be difficult to reconcile with the DM interpretation of the GC TeV flux.
Finally, since the dark matter permeates our Universe, if the same radiation
was found in the centers of other mass
concentrations, population studies may be possible that could help confirm or
deny the annihilation
nature of this radiation \cite{vassiliev02}.
\par
A small spike on an NFW profile could explain the large gamma-ray flux which
is not expected from cored or cusped halos.
Astrophysically small spikes in the DM halos are not favored, but not ruled
out either.  The infrared data of proper motions in the GC show about
three million solar masses confined to a space of 90 AU.  
The compression of this baryonic matter may adiabatically
compress the dark matter and lead to such a spike in the
profile \cite{gondolo99}.  Any merger events with larger stellar sized objects
should dynamically heat the DM spike reducing its density.  
\par
Further observations of the Galactic Center in gamma rays are ongoing.
There are hints that the TeV radiation from the Galactic Ridge is connected to
the Galactic Center point source.
The TeV flux from the GC seems to be constant
in time and a cutoff in the spectrum (now reported to have a spectrum with
$\alpha = 2.4$) has not been found up to energies of
$\sim6$ TeV \cite{rolland05}, so the models considered here are still viable.
The nature of this non-thermal radiation source in the center of the Milky Way
is still unknown and undergoing active study and observations.
\vspace{2\baselineskip}

{\it Acknowledgements.} This work was supported in part by NSF grant
PHY-0456825 and NSF grant \#0079704. We thank Mark Morris for useful
discussions and Ted Baltz for providing us with a collection of minimal
supergravity models with heavy neutralinos.

\end{document}